\documentclass[aps,twocolumn,superscriptaddress,preprintnumbers]{revtex4}

\usepackage{graphicx,psfrag}
\usepackage{amsfonts}
\usepackage[figuresright]{rotating}  
\usepackage{amssymb}
\usepackage{amsmath}
\usepackage{subfigure}
\usepackage{multirow}
\usepackage{tabularx}

\def\avg#1{\langle#1\rangle}

\def\beq{\begin{eqnarray}}
\def\eeq{\end{eqnarray}}

\begin{document}
\title{Topological semimetal in a fermionic optical lattice}

\author{Kai Sun}
\affiliation{Condensed Matter Theory Center and Joint Quantum Institute,
Department of Physics, University of Maryland, College Park, MD 20742, USA}
 
\author{W. Vincent Liu \footnote{email: w.vincent.liu@gmail.com.}}
\affiliation{Department of Physics and Astronomy, University of
  Pittsburgh, Pittsburgh, PA 15260, USA}
\affiliation{Kavli Institute for Theoretical Physics, University
of California, Santa Barbara,  CA 93106, USA}
\affiliation{Center for Cold Atom Physics, Chinese Academy of
  Sciences, Wuhan 430071, China}

\author{Andreas Hemmerich}
\affiliation{Institut f\"ur Laser-Physik, Universit\"{a}t Hamburg, Luruper Chaussee 149, 22761 Hamburg, Germany}

\author{S. Das Sarma}
\affiliation{Condensed Matter Theory Center and Joint Quantum Institute, 
Department of Physics, University of Maryland, College Park, MD 20742, USA}
\preprint{NSF-KITP-10-148}
\maketitle

\textbf{Optical lattices play a versatile role in advancing our understanding 
 of correlated quantum matter.  
 The recent implementation of orbital degrees of freedom 
 in chequerboard~\cite{Wirth2010,Olschlager2010} and hexagonal~\cite{Sengstock2010} 
 optical lattices opens up a new thrust
 towards discovering novel quantum states of matter, which have no prior
 analogs in solid state electronic materials. Here, we demonstrate that an exotic
 topological semimetal emerges as a parity-protected gapless state
 in the  orbital bands of a two-dimensional fermionic optical lattice.  The new quantum
 state is characterized by a parabolic band-degeneracy point with
 Berry flux $2\pi$, in sharp contrast to the $\pi$ flux of Dirac
 points as in graphene.  We prove that the appearance of this
 topological liquid is universal 
 for all lattices with D$_4$ point group symmetry as long as orbitals with
 opposite parities hybridize strongly with each other and the band
 degeneracy is protected by odd parity. Turning on inter-particle
 repulsive interactions, the system undergoes a phase transition to a
 topological insulator whose experimental signature includes chiral
 gapless domain-wall modes, reminiscent of quantum Hall edge states. }

The search for topological states of matter has been a focus of
theoretical and experimental studies, since the discovery of the
quantum Hall effect (See the review of
Ref.~\cite{Nayak2008}
and references therein). This problem was
brought to the forefront again recently by the theoretical prediction and
experimental discovery of the time-reversal invariant Z$_2$
topological insulators in semiconductors with strong spin-orbit
couplings~\cite{Kane2005, Bernevig2006, Konig2007, Fu2007,Moore2007,Roy2009,Hsieh2008}.
(For more details see the recent reviews of Refs. \cite{Hasan2010,Qi2010} and
references therein).
For noninteracting particles, the topological properties of insulators 
as well as topological superconductors have recently been classified based on 
the anti-unitary symmetries of the systems~\cite{Kitaev2009,Schnyder2008}.
However, this elegant topological classification does not apply to Fermi
liquid (metal or semimetal) states due to the existence of fermionic
low-energy modes in gapless systems.  
In this paper, we shall show, however, that a novel type of
topologically-nontrivial  
semimetal  unexpectedly arises as a universality class
for arbitrary two-dimensional lattices with D$_4$ point group symmetry due to
the mixing of orbitals of opposite parity. We believe that our discovery should be 
realizable in fermionic cold atom optical lattices rather easily.

\begin{figure}[htp]
\begin{center}
\includegraphics[width=0.43\textwidth]{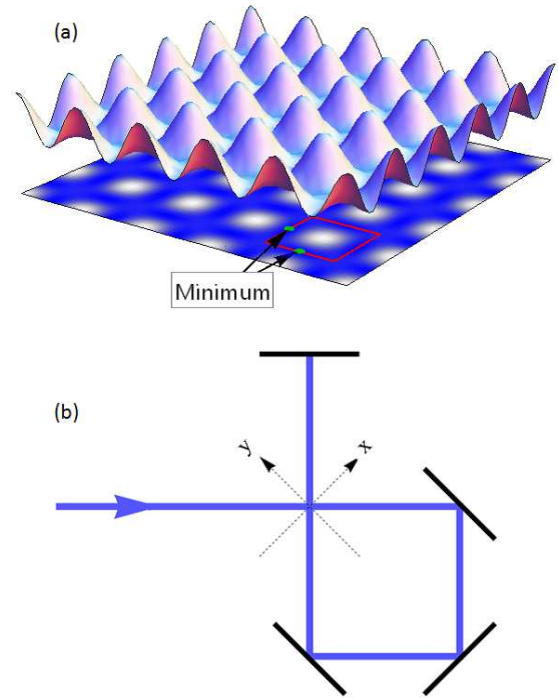}
\end{center}
\caption{(a) The optical lattice  potential in equation~\eqref{eq:lattice}. 
Here we choose $V_1=2.4 E_R$
and $V_2=1.6 E_R$, where $E_R=h^2/(2 m \lambda^2)=h^2/(4ma^2)$ is the recoil energy with $h$ being the Planck constant, $m$ being the mass of the particle, $\lambda$ being the wavelength of the light beam and  $a$ being the lattice constant.
The plane at the bottom shows the contour plot of the same potential. The red square marks an unit cell and the green dots indicate the two 
energy-minimum points of this unit cell located at the bond centers.
(b) The experimental setup to realize the lattice potential in equation~\eqref{eq:lattice} 
for $V_2/V_1 \geq 1/2$. 
The linear polarization of the incident monochromatic light beam (solid blue line) encloses an 
angle $\alpha$ with respect to the normal direction to the drawing plane. The black bars represent mirrors and the dashed arrows mark the $x$ and $y$ directions of the coordinates.  
See Methods for analysis.
}
\label{fig:lattice}
\end{figure}  

\begin{figure}[htp]
\begin{center}
\subfigure[]{\includegraphics[width=0.23\textwidth]{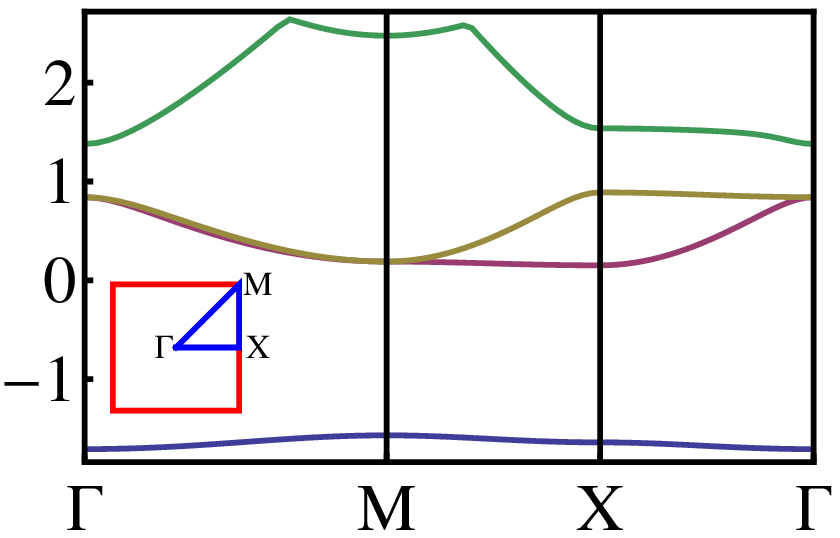}}
\subfigure[]{\includegraphics[width=0.23\textwidth]{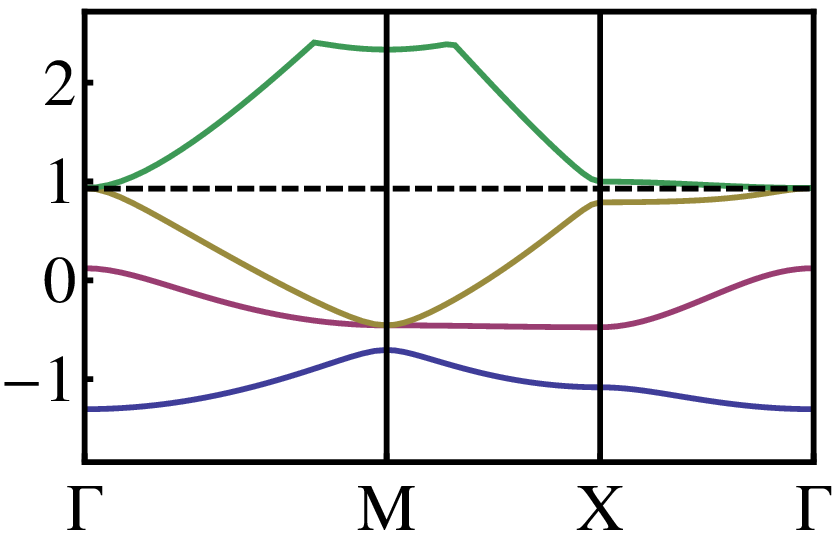}}
\end{center}
\caption{The single-particle energy spectrum  (measured in the unit of $E_R$) 
for the lowest four bands and the topological structure near band degeneracy points. 
(a) and (b) show the band structure for the momenta along the contour
from $\Gamma$ to $M$ to $X$ and back to $\Gamma$. This contour is
shown in the inset in (a) represented by the blue lines, while the red square marks the
Brillouin zone. At $V_1=2.4 E_R$, 
two different types of band structures are observed. (a) shows the band structure
at $V_2<0.87 E_R$ ($V_2=0.4 E_R$), where the hybridization between different orbitals is 
weak. We refer to this type of band structure as the weak-hybridization limit. 
The band structure at larger $V_2$ was shown in (b) with $V_2=1.6 E_R$. This case is 
referred to as the strong-hybridization limit. The dashed line in (b) marks the chemical potential, 
at which the system becomes a topological semimetal. The marginal case 
$V_2=0.87 E_R$ is shown in the Supplementary Information, where all the three upper bands touch at $\Gamma$ point.}
\label{fig:band_structure}
\end{figure} 

The physics of higher orbtials in optical lattices has recently
emerged as an exciting new front in both
theoretical~\cite{Girvin:05,Liu-Wu:06,Kuklov:06} and experimental
(e.g., early \cite{Kohl2005,Browaeys2005,Lee2007,Muller2007} and recent
\cite{Wirth2010,Olschlager2010,Sengstock2010}) 
studies.
We specifically examine a model system that resembles the
D$_4$ symmetric double-well lattice reported by the Hamburg
experimental group~\cite{Wirth2010,Olschlager2010}, but our conclusions
apply generally to other lattices with the same point group symmetry.
Consider the optical lattice shown in Fig.~\ref{fig:lattice}a with the potential
\beq
V(x,y)&=&-V_1 [\cos (k x)+\cos (k y)] \nonumber\\
&&+ V_2[ \cos (k x+ k y)+\cos (k x - k y)].
\label{eq:lattice}
\eeq
Here, $k=2\pi/a$ and $a$ is the lattice constant. $x$ and $y$ are the 2D coordinates in 
configuration space.
The parameters $V_1$ and $V_2$ are chosen to be positive. This optical lattice can be formed using a single chromatic light field following the experimental 
setup shown in Fig.~\ref{fig:lattice}b for $V_2/V_1\ge 1/2$. 
For completeness, below we will first consider the general situation 
with $V_2/V_1\ge 0$. Then, we will show that  the parameter range
of interest in our work is $V_2/V_1\sim 2/3 >1/2$, which can be realized
using the proposed experimental setup shown in Fig. 1b and discussed below 
in the Methods section.

For $V_2=0$, the $V_1$ term induces a square lattice with lattice constant $a$. 
As $V_2$ increases, the potential energy at the center of a unit 
cell [with coordinates $(0,0)$] is increased while the potentials near the bond centers [with coordinates $(\pm a/2,0)$ and $(0,\pm a/2)$] are reduced. For $V_2>V_1/2$,  each unit cell contains 
potential minima located at $(\pm a/2,0)$ and $(0,\pm a/2)$ as shown in Fig.~\ref{fig:lattice}a.

We numerically solve the band structure of this lattice via plan-wave
expansions and find that band degeneracy points appear between higher orbital bands
at $\Gamma$ and $M$ points (the center and corner of the Brillouin
zone). For the lowest four bands, 
as shown in Fig.~\ref{fig:band_structure}, in the small $V_2$ limit, the second
and third bands cross at both $\Gamma$ and $M$ points. For larger
$V_2$, there are still two band degeneracy points for the lowest four
bands, but now the second and third bands only cross at $M$,
while the third and fourth bands become degenerate at $\Gamma$. 
For even larger $V_2$ (not shown), the first and second bands
become degenerate at $M$, while the third and fourth bands touch at
$\Gamma$. This large $V_2$ limit is dominated by the same physics as 
in the intermediate $V_2$ regime, and thus, we will only focus on the small
and intermediate $V_2$ in this paper.

The band degeneracy phenomenon described above is generic and stable. In fact, as shown 
in the Supplementary Information, for noninteracting particles, these band degeneracy 
points are topologically protected and remain stable when system 
parameters are tuned 
adiabatically, as long as the lattice point group symmetry is maintained (although a band 
degeneracy point may move from between the $n$ and $n+1$ bands to the $m$ and 
$m+1$ bands as shown in the examples above).  
With details presented in Methods and the Supplementary Information, 
near the band degeneracy point, a 2D vector field ($\vec{h}_{\vec{k}}$) in momentum 
space can be defined using the Hamiltonian of the system. At the momentum $\vec{k}$, 
the length of this 2D vector ($|\vec{h}_{\vec{k}}|$) gives (half of) the energy splitting 
between two energy bands. 
For the band degeneracy points in our model, this vector field possesses a topological 
defect, a vortex with winding number $2$. 
At the vortex core, the length of the vector vanishes ($|\vec{h}_{\vec{k}}|=0$), indicating that the band gap vanishes here (i.e., a band degeneracy point). 
It is this topological property that dictates the stability of the band degeneracy against any adiabatic deformation. From the mathematical point of view, this nontrivial topology can be described rigorously using the topological index of the Berry flux, which is $2\pi$ for this case.

In addition, the band degeneracy point is also protected by
the parity of the Bloch wavefunctions under space inversion.
In fact, as shown in the Methods section, 
it turns out that all the essential physics of the topological semimetal
can be understood
within a simple tight-binding picture without considering the full band structure theory,
and the key ingredient for this phenomenon is the mixing between the orbitals of opposite parity. 
In the particular model we consider here, the semimetal is formed by
the hybridization between the $d$ orbital and the two $p$ ($p_x$ and
$p_y$) orbitals at each lattice site. 

We now study the instability of the topological semimetal in
the presence of interaction, with details presented in Supplementary Information. 
We start 
with the tight-binding Hamiltonian and derive an effective low-energy theory
around the Fermi point, which in this case is the degeneracy point of the 
third and fourth
band (Fig.~\ref{fig:band_structure}b). It turns out that this
effective theory in the presence of interaction can be
mapped onto a general theoretical
model of $d$-wave symmetry which was analyzed in 
Refs.~\cite{sun2009,sun2008} via the renormalization group technique. 
Therefore, by mapping the results back from that $d$-wave
model, we obtain the universal property for the band degeneracy point
of the topological semimetal we present here. Below, we summarize the main
results.

As temperature is lowered below a critical
value, $T_c$, the system undergoes a second order phase transition,
where $T_c\sim W e^{-\alpha/N(0) V}$ with $N(0)$ the density of
states at the chemical potential, $V$ the interaction strength
and $W$ the band width. The parameter $\alpha$ 
is a dimensionless constant whose value is determined by the band structure.
In our model, the order parameter describing this 
low-temperature ordered phase is the $z$-component of the angular 
momentum  $\avg{L^{z}_{\vec{r}}} =-i \avg{p_{x,\vec{r}}^\dagger 
p_{y,\vec{r}}-p_{y,\vec{r}}^\dagger p_{x,\vec{r}}}$, where $p_{x,\vec{r}}$
and $p_{y,\vec{r}}$ are the fermion annihilation operators of the $p_x$ and $p_y$
orbitals on site $\vec{r}$.  (This order parameter can be mapped to the order 
parameter $\Phi$ in the general theory studied in Ref~\cite{sun2009}.)

In our system, the repulsive interaction can be reformulated as
\begin{align}
H_{\textrm{int}}=V \sum_{\vec{r}} p^\dagger_{x,\vec{r}} p_{x,\vec{r}} p^\dagger_{y,\vec{r}} p_{y,\vec{r}} 
=-\frac{V}{2} \sum_{\vec{r}} (L^{z}_{\vec{r}})^2,
\label{eq:interactions}
\end{align}
where $V>0$ is the interaction strength. This interaction term favors a state with nonzero
angular-momentum $\avg{L^z}\ne 0$. In an ordinary metal or insulator (or
graphene \cite{DasSarma2009}), the formation of nonzero angular momentum costs kinetic
energy which usually dominates over the energy gain from
interaction unless the interaction strength is very large.
However, for the topological semimetal we find here, the energy cost for nonzero
$L^{z}$ from the kinetic part is always subleading compared with the energy gain
from interaction at low enough temperature. This results in the
spontaneous generation of angular momentum, which is a key theoretical insight of our work.

From the symmetry point of view, this low temperature phase
spontaneously breaks the D$_4$ point group symmetry down to C$_4$, and
also breaks the time-reversal symmetry. This symmetry breaking pattern
belongs to the Ising universality class resulting in two degenerate
ground states with opposite angular momentum.

\begin{figure}
\begin{center}
\includegraphics[width=0.4\textwidth]{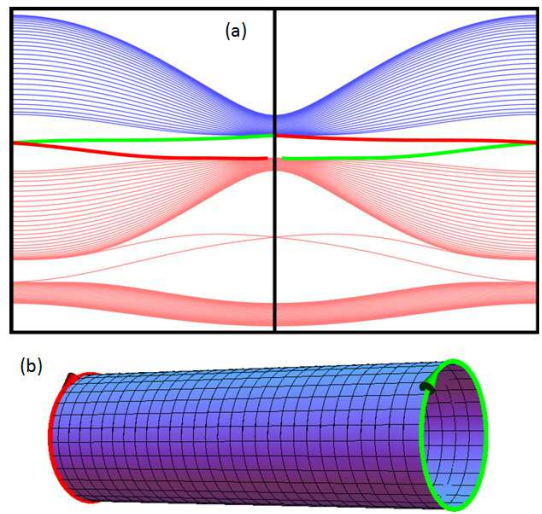}
\end{center}
\caption{Topologically protected edge states and domain-wall
 modes. Figure (a) shows 
 the single-particle energy spectrum of the insulating phase with
 $\avg{L^z}\ne 0$ computed within mean-field
approximation on a cylinder geometry (see Supplementary Information for technical details). 
The horizontal axis is the momentum defined along the periodical direction of the cylinder (from $-\pi/a$ to $\pi/a$) and the vertical axis is the energy. The pink curves at the bottom of (a) describes the states in the valence bands filled by  particles, while the blue curves on the top part are the empty band. The green and red curves are the chiral gapless edge states located on the two edges of the cylinder as shown in (b). Here, figure (b) is a schematic picture showing the geometry of the system we used to compute the edge states. The black solid lines show the underlying square lattice. 
The two thick lines at the edges (red and green) represent the chiral edge states, with arrows indicating the chirality. The length of the cylinder we used is $30 a$ with $a$ being the lattice spacing. 
In this case, the finite-size effects are negligibly small.
}
\label{fig:edge_states}
\end{figure}

As for the band structure, the band degeneracy at $\Gamma$ is lifted 
in the symmetry broken phase. (The degeneracy at $M$ is also lifted. However, 
this is not relevant to our study since that degeneracy point is located far below 
the chemical potential.) Hence the topological semimetal becomes a fully gapped 
insulator in the presence of interaction with the gap being $V \avg{L^{z}}$. 

This insulator turns out to be topologically nontrivial characterized by the nontrivial value of the topological index, known as the first Chern number. The Chern number for this state is $1$, which indicates that this system shares the same topological properties as the quantum Hall state with filling $1$. However, in contrast to the quantum Hall effect where the nontrivial topological state 
is induced by a strong external magnetic field, here the same quantum topological state of matter originates from many-body effects in the absence of any external magnetic field. In general, states
with nonzero Chern number in the absence of an external magnetic field are
known as the anomalous quantum effect states, first proposed in a toy
model on a honeycomb lattice by Haldane~\cite{Haldane1988}. Recently,
several different possible realizations of the Haldane model in cold
gases were discussed using lattice rotations~\cite{Wu2008} or
light-induced vector potentials~\cite{Stanescu2010}. In our predicted
topological phase, however, interaction plays a decisive role  in
sharp contrast to the noninteracting situation prevailing in the
quantum Hall effect or anomalous quantum Hall effect. To the best of
our knowledge, our work is the only theoretical prediction 
in the literature of an interaction-driven anomalous quantum Hall state.

Furthermore, if two spin components are both present in the atomic gases, the same interaction effect may lead to a time-reversal  invariant Z$_2$ topological insulator. This phenomenon can be partially understood as an interaction-driven 2D-version of HgTe. As pointed out in Ref.~\cite{Fu2007},
the combined effect of spin-orbit coupling and strain opens a gap at a 3D quadratic band degeneracy point and leads to a 3D topological insulator.
By contrast, in our 2D system, topological states arise purely due to many-body interaction effects.

To further demonstrate the topological nature of this insulating phase, we computed the band structure of this state on a cylinder, as shown in Fig.~\ref{fig:edge_states}. Here, although the bulk modes are all gapped, there is a gapless topological chiral edge state on each of the two edges of the system, which is the direct signature of a topologically nontrivial insulator.

The phase transition being discussed in our work has strong analogy to the BCS theory of superconductivity. In
particular, the two classes have similar scaling formula for the mean-field transition temperature ($T_C\sim e^{\alpha /N(0) V}$). However,  the phase transition here breaks only a discrete symmetry (time-reversal) and thus belongs to the Ising universality class. In 2D, the fluctuation effect is weak for an Ising transition
and long-range order is sustained at finite temperature. On the contrary, the BCS transition breaks the continuous $U(1)$ symmetry and belongs to the $XY$ universality class. As a result, the BCS transition in 2D is a Kosterlitz-Thouless transition, whose transition temperature is strongly suppressed by phase fluctuations and is much lower than the mean-field prediction. Thus the transition temperature for our problem should be much higher than the BCS transition, if all 
other parameters ($N(0)$, $V$, etc.) have the same value.
Therefore, under equivalent conditions the phase transition predicted by us should be much easier to
observe in 2D optical lattices than the corresponding BCS Kosterlitz-Thouless transition.

Beyond its theoretical significance, the topological semimetal state also 
has robust and unique experimental signatures. For example, the energy band structure of the 
unique band-crossing degeneracy point can be detected  directly using experimental techniques, 
such as Bragg scattering~\cite{Ernst2010} as discussed in our Supplementary Information.

At low temperatures, the system remains a topological semimetal for attractive
interactions but becomes an insulator for repulsive interactions. Since both the values
and the signs of interaction can be tuned in ultra-cold gases, this phase 
transition, between a compressible liquid and an incompressible insulator, can be studied 
experimentally by measuring the compressibility at different interactions. In addition, Bragg 
scattering can also be used to detect the insulating gap induced by the repulsive interactions. 
Because the low-temperature topological insulating state spontaneously breaks
 the time-reversal symmetry, any experimental
measurements sensitive to the time-reversal symmetry can also be used to
identify this phase, such as the Hall effect.

The direct experimental evidence for a topological insulator is the
gapless chiral edge state which is a metallic state localized on the edge of 
a topologically-nontrivial insulator. However, it is worthwhile to note that
the sharp edge in the
condensed matter system is absent in cold atomic gases. Due to the
existence of the slowly varying trap potential, one expects the density to decrease
away from the center of the trap. Therefore, the system is a
liquid near the edge due to the low filling fraction. This
liquid state from incommensurate filling will hybridize with the
topological edge state, which makes the observation of the topological
edge states challenging in atomic systems. This difficulty can be avoided if two domains
of topological insulating phases with opposite angular momenta are
induced. At the domain wall between these two areas, 
compressible chiral domain-wall states should exist. Since this domain wall can be 
chosen to locate near the center of the trap, far away from the trivial liquid state near the 
edge of the system, it should in principle provide a clean signature for the
topological edge states. 
 These domain-wall modes can also be detected using Bragg scattering, where 
one finds that the insulating gap is reduced to zero near the domain-wall. In 
each real experimental system, due to finite number of particles on 
a particular optical lattice,
the vanishing of the insulating gap at the domain wall is in fact prohibited 
by finite size effects.
For topological insulators, such finite size effects have been systematically studied and
the metallic edge states are found to be detectable even for a system with about 10 particles~\cite{Varney2010}.
An alternative experimental way of seeing the topological edge state 
would be to have a sharp trap boundary, as in a square-well potential, which would suppress 
the hybridization between the trivial liquid phase and the edge topological state.
In such a square-well trap, the topological edge state should manifest itself directly.

\section*{Methods}
\paragraph*{\bf Creation of the optical lattice.}
In the experimental setup shown in Fig.	~\ref{fig:lattice}b of the Letter, 
by superimposing two monochromatic optical standing waves oscillating in phase, 
we implement the electric field
\begin{align}
E=&
\epsilon
\begin{pmatrix}
-\frac{1}{\sqrt{2}} \sin\alpha
\\
\frac{1}{\sqrt{2}} \sin\alpha
\\
\cos\alpha
\end{pmatrix}\cos\left[k (x+y)/2\right]
\nonumber\\
&-\epsilon
\begin{pmatrix}
\frac{1}{\sqrt{2}}\sin\alpha
\\
\frac{1}{\sqrt{2}}\sin\alpha
\\
\cos\alpha
\end{pmatrix}\cos \left[k (x-y)/2\right].
\end{align}
The corresponding light shift potential is $U(x,y)=-\chi|E(x,y)|^2$
with $\chi$ denoting the  real part of the polarizability.  
It is straight froward to check that this potential is identical to the potential we proposed in the
letter, up to a trivial constant:
\begin{align}
U(x,y)=&-V_1 [\cos (k x)+\cos (k y)]
\nonumber\\&
+ V_2[ \cos (k x+ k y)+\cos (k x - k y)]-\chi \epsilon^2,
\end{align}
with  
\begin{align}
V_1=&-\chi\epsilon^2\cos^2\alpha,
\\
V_2=&-\chi \epsilon^2/2.
\end{align}
By choosing blue detuning, i.e., $ \chi<0$, we obtain $V_1>0$ and $V_2 >0$. When the polarization direction,  $\alpha$, is changed, the ratio $V_2/V_1$ can be tuned to any value above $1/2$. 
For example, using fermionic potassium $^{40}$K with a principle fluorescence line at $767$nm, 
a standard green frequency-doubled Nd:YAG-laser (532 nm) would be a suitable light source for implementing the desired optical potential.

\paragraph*{\bf Hybridization between orbitals of opposite parity.}
It turns out that all the essential physics of the topological
semimetal can be understood within a simple tight-binding picture and
the key-ingredient for this phenomenon is the mixing between orbitals
with opposite parities under space inversion.  
Here, we outline the main procedures and results of calculation for
the mixing of parity even $d_{x^2-y^2}$ and odd $p_x$ and $p_y$
orbitals, and defer the details (e.g., model Hamiltonian, band
structure, etc.) to the Supplementary Information (Section S-6). 
In this study, these three orbital bands are considered next to the
chemical potential level and 
all other orbitals are assumed to be far separate from them (such that their effects
can be dynamically ignored).  When the mixing between the
two types of orbitals is weak, the parity-odd orbitals form two bands,
which cross each other at the $\Gamma$ and $M$ points, while the band
formed by the parity-even orbitals show no degeneracy (similar to
Fig.~\ref{fig:band_structure}a which was obtained by numerical
diagonalization).   In contrast, as the mixing between different
types of orbitals is enhanced, the three bands formed by these three
orbitals hybridize together, and now the middle band crosses with both
the other two bands, one at $\Gamma$ and another at $M$, similar to
Fig.~\ref{fig:band_structure}b. In fact, the top three bands shown in
Fig.~\ref{fig:band_structure} are mainly contributed by the $p_x$,
$p_y$ and $d$ orbitals. In the Supplementary
Information, a full comparison is provided between the band structure
of the optical lattice model defined by the potential
\eqref{eq:lattice} and that of the effective three-orbital ($p_x,p_y,
d_{x^2-y^2}$) tight-binding model.

\paragraph*{\bf Instability under infinitesimal repulsion.}
Using the conclusions from 
Ref.~\cite{sun2009}, we found that under renormalization group, 
the repulsive interaction shown in equation~\eqref{eq:interactions} 
is a marginally-relevant perturbation and it is also the only 
relevant perturbation for spinless fermions with short-range interactions. 
Therefore, at low temperature, this interaction term dominates the low-energy 
physics and will stabilize a state with nonzero angular momentum $\avg{L^z}\ne 0$.
This state is a topological insulator with Chern number $1$,
in agreement with the general study shown in Ref.~\cite{sun2009}.
This conclusion is further verified in Fig.~\ref{fig:edge_states}, where we 
examined the mean-field single-particle spectrum for a cylindric 
geometry and observed the gapless chiral edge states. 

\section*{Acknowledgment}
We appreciate the very helpful discussions with L. Fu, C. L. Kane and X.-L. Qi. The work of K. S. and S. D. S. is supported by JQI-NSF-PFC, AFOSR-MURI, ARO-DARPA-OLE, and ARO-MURI.  W.V.L. is supported by ARO (W911NF-07-1-0293) and ARO-DARPA-OLE (W911NF-07-1-0464). We thank the Kavli Institute for Theoretical Physics at UCSB for its hospitality where this research is supported in part by National Science Foundation Grant No. PHY05-51164. 

\section*{Author Contributions}
WVL, KS, and SDS planned the work.  KS and WVL carried out most of the calculations 
with input from SDS. AH provided the experimental protocol. All authors 
contributed to the writing of the manuscript.

\section*{Author Information}
The authors declare no competing financial interests.
Correspondence and requests for material should be sent to w.vincent.liu@gmail.com.
Supplementary information accompanies this paper.

\onecolumngrid

\renewcommand{\thesection}{S-\arabic{section}}
\renewcommand{\theequation}{S\arabic{equation}}
\setcounter{equation}{0}  
\renewcommand{\thefigure}{S\arabic{figure}}
\setcounter{figure}{0}  

\section*{\Large\bf Supplementary Information}

\section{Optical lattice}
\begin{figure}[htp]
\begin{center}
\subfigure[]{\includegraphics[width=0.3\textwidth]{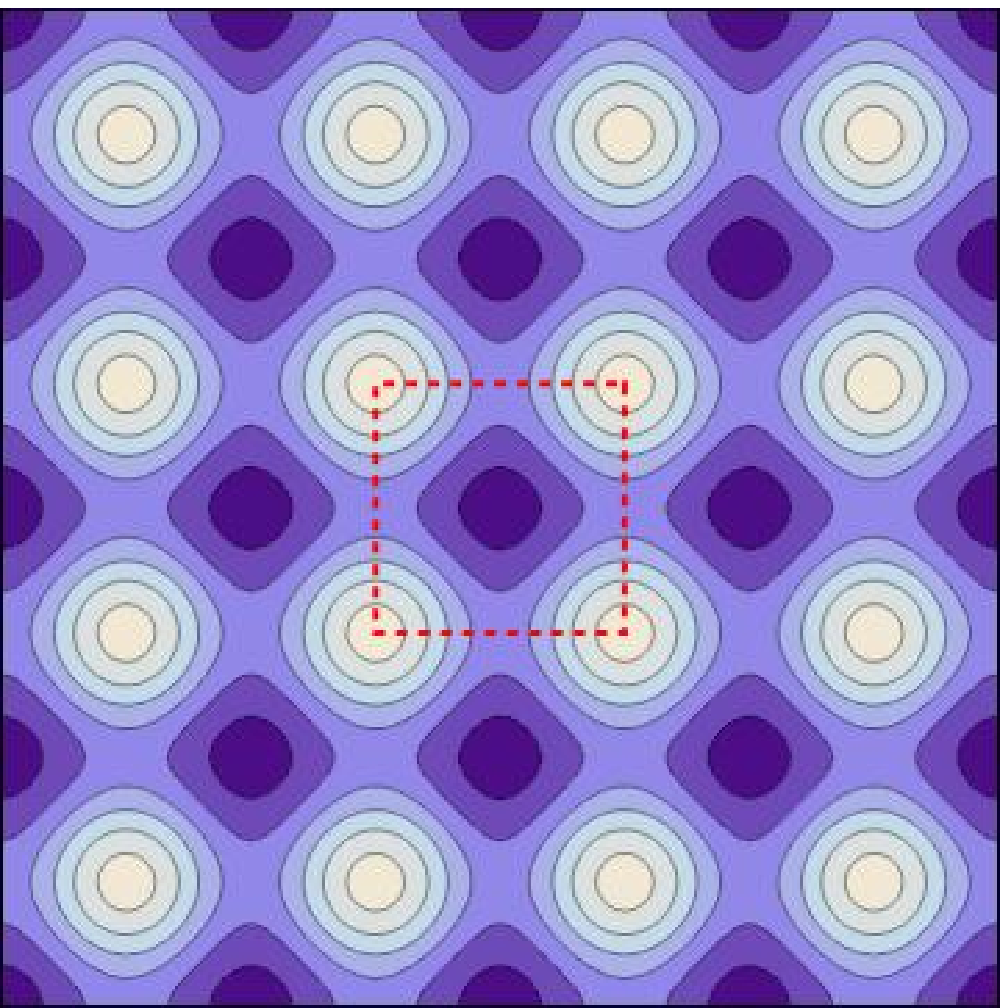}}
\subfigure[]{\includegraphics[width=0.3\textwidth]{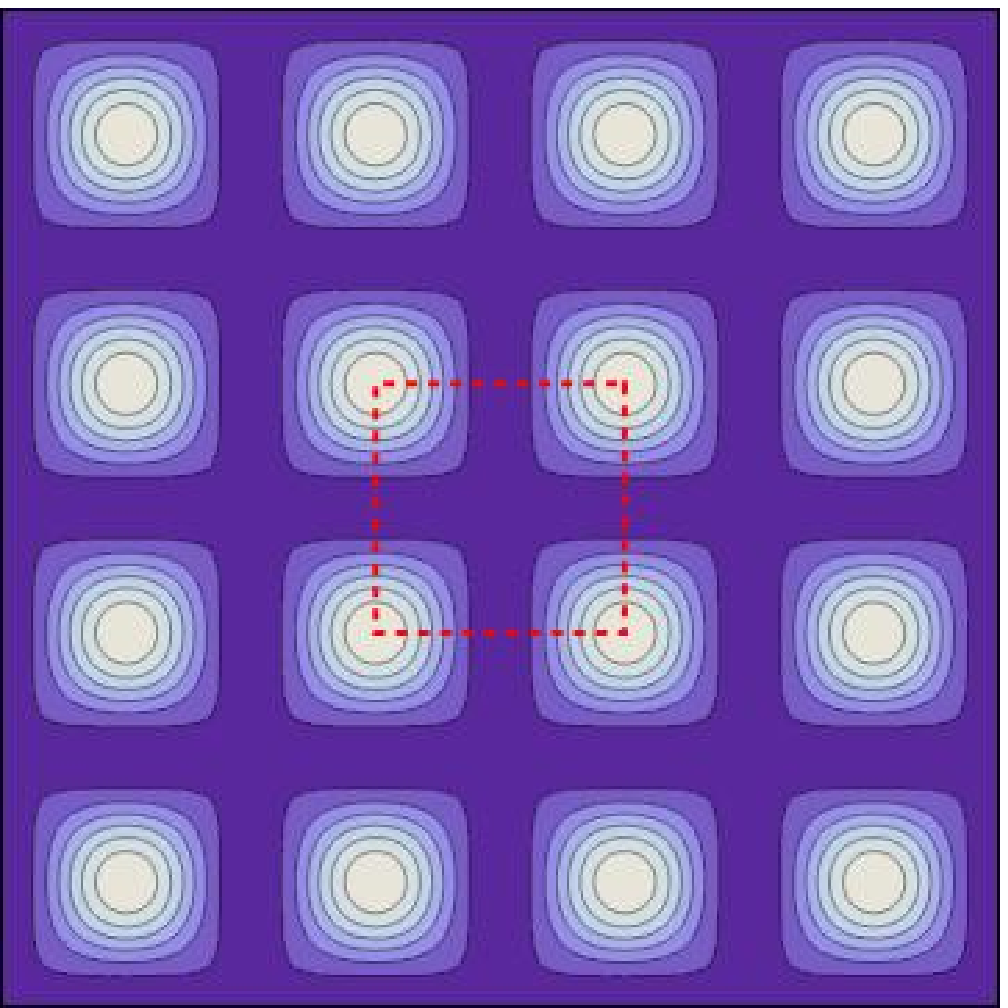}}
\subfigure[]{\includegraphics[width=0.3\textwidth]{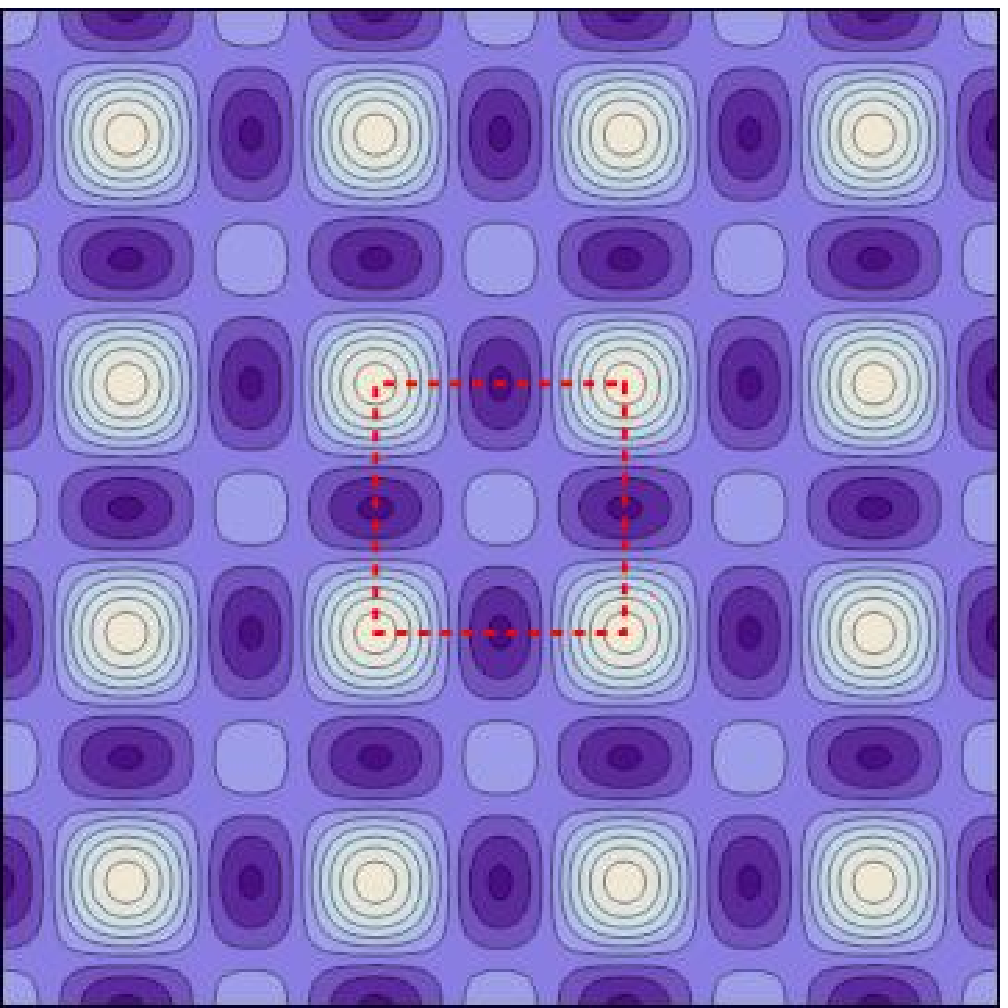}}
\end{center}
\caption{The optical lattice shown in equation~\eqref{app:eq:lattice} at different parameters.
The darker (lighter) regions represent areas where the potential is low (high). The dashed line
marks one unit cell of the lattice. In (a-c),
$V_1=2.4 E_R$ and $V_2=0.4 E_R$, $1.6 E_R$ and $3.2 E_R$.}
\label{fig:lattices}
\end{figure} 

Figure \ref{fig:lattices} shows the optical lattice described by
\begin{align}
V(x,y)=&-V_1 [\cos (k x)+\cos (k y)]
+ V_2[ \cos (k x+ k y)+\cos (k x - k y)].
\label{app:eq:lattice}
\end{align}
In Fig.~\ref{fig:lattices}b, we employed the same parameters used in the letter to derive the topological semimetal.
For $V_2<V_1/2$ [Fig. \ref{fig:lattices}.(a)], the optical lattice is basically a simple square lattice with a single energy minimum in each unit cell located at the center. However, when $V_2> V_1/2$ [Fig. \ref{fig:lattices}.(b) and (c)], there are two energy minima in a unit cell, located at the bond-center. 
As for the marginal case $V_2=V_1/2$ (not shown), the minimum of the potential is reached when $x$ or $y$ becomes $2 n\pi/k$ with $n$ being any integer.

\section{Plane-wave expansions and numerical calculation of the band structure}

We expand the Bloch wavefunction in the basis of plane waves
\begin{align}
\Psi^n_{\vec{k}}(\vec{r})=\sum_{n,m} a_{n,m}e^{i (2\pi n /a+k_x) x+i (2\pi m /a+k_y) y},
\label{eq:plan_wave_expansion}
\end{align}
where $a_{n,m}$ are the complex coefficients with $n$ and $m$ being integers and 
$\vec{r}=(x,y)$ is the real-space coordinates. In the lattice Hamiltonian, the potential
energy introduces hybridization between the plane-waves with different $n$  and $m$.
For the low-energy bands, the contributions from plane waves with large $m$ and $n$ are small. 
Thus, we ignore states with $n$ and $m$ larger than certain cutoff $N$. In the numerical calculation demonstrated here, $N\sim20$. And the band
structure is independent of the choice of $N$ for $N>10$ (up to negligible small corrections to the eigenenergy.)

\begin{figure}
\begin{center}
\subfigure[]{\includegraphics[width=0.3\textwidth]{band1}}
\subfigure[]{\includegraphics[width=0.3\textwidth]{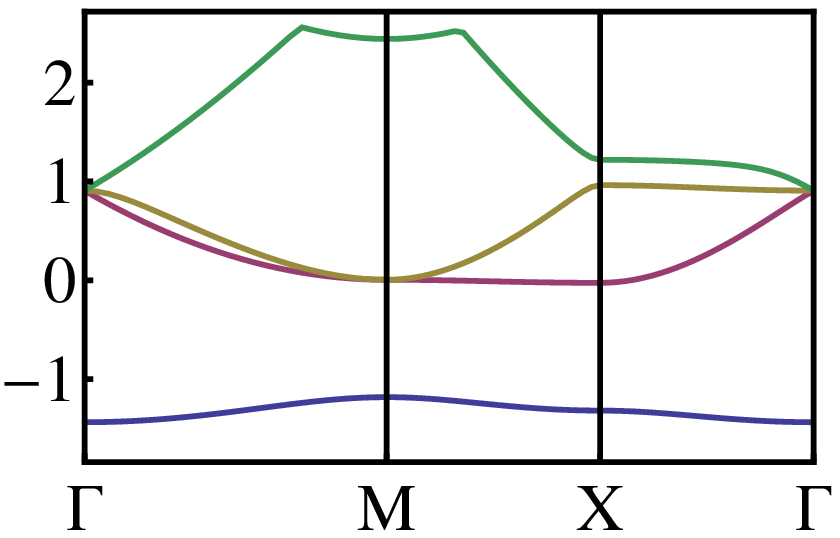}}
\subfigure[]{\includegraphics[width=0.3\textwidth]{band2}}
\end{center}
\caption{The single-particle energy spectrum of the lowest four bands.
Here, we show the band structure for the momenta along the contour from $\Gamma$ to $M$ to $K$ and back to $\Gamma$, as shown in the inset of Fig.~(a). 
The parameters here are chosen to be $V_1=2.4 E_R$ while 
$V_2$ takes the value of (a) $0.4 E_R$, (b) $0.87 E_R$, and (c) $1.6 E_R$, respectively.}
\label{fig:bands}
\end{figure}

Using this approach, the Hamiltonian is a $(2 N+1)^2\times(2N+1)^2$
matrix. We find the eigenvalue and eigenvector of this matrix at each momentum in the Brillouin zone.
The eigenvalues (as a function of $\vec{k}$) provides the single-particle band structure,
while the eigenvector gives the coefficients of $a_{n,m}$ in equation~\eqref{eq:plan_wave_expansion}.
In Fig.~\ref{fig:bands}, the band structure is shown at $V_1=2.4 E_R$ for different values of $V_2$.

\section{Degeneracy of parity odd orbitals in a potential well with four-fold rotational symmetry}
\label{sec:single_site}

Before studying the band degeneracy on a lattice, we examine the degeneracy of orbitals in a single
potential well. The conclusions discovered here shall be generalized to the study of band degeneracy in the next section.

Consider a particle trapped in a 2D potential well with four-fold rotational symmetry.
The symmetry group describing this rotational symmetry is the dihedral group of D$_4$.
We define $\phi^m(\vec{r})$ with $m=1,2,\ldots$ to be the eigen-wavefunctions.
Under a certain rotation in the symmetry group, $R \in \textrm{D}_4$, 
the wavefunctions $\phi^m(\vec{r})$ reads
\begin{align}
\phi^n(\vec{r}) =\sum_m U_{n,m}(R) \phi^m(\vec{r}),
\end{align}
where $U_{n,m}(R)$ is the $(n,m)$ element of the unitary matrix $U(R)$. 
By definition, the matrices $U(R)$ for all rotations in the point group forms a 
unitary representation of the symmetry group D$_4$.

For any unitary representation of D$_4$, it can always be separated into the direct sum of irreducible
representations. In terms of the matrix $U(R)$, this means that it can be block diagonalized. Each block
corresponds to an irreducible representation of D$_4$. For a n-dimensional  irreducible representation, 
the block has the size of $n\times n$.

In general, the states within a block (which belongs to the same irreducible representation of D$_4$) 
are related to each other via rotation. Since the system is invariant under D$_4$, these states  must have the same energy. Thus, for a state that belongs to an irreducible representation of dimensions $n>1$, it must be $n$-fold degenerate.

For the D$_4$ group, there are 4 one-dimensional representations 
(known as $A_1$, $A_2$, $B_1$ and $B_2$) and 1 two-dimensional one ($E$).
Interestingly, all the one-dimensional representations of D$_4$ are even under space-inversion
\begin{align}
\phi^m(\vec{r})=\phi^m(-\vec{r}).
\end{align}
On the other hand, the two-dimensional representation $E$ has an odd parity
\begin{align}
\phi^m(\vec{r})=-\phi^m(-\vec{r}).
\end{align}
Therefore, we conclude that any space-inversion odd state in this system is two-fold
degenerate, while space-inversion even states are in general nondegenerate.

\section{Space-group symmetry and band degeneracy points}
\label{sec:space_group}
The crystal symmetries of a lattice are described by the corresponding space group 
$G$. Each element of the space group $\hat{g}=\{R|\vec{v}\}$ is a symmetry operation which combines a space rotation $R$ (represented by a $d\times d$ orthogonal matrix with $d$ being the
spatial dimensions of the system) and a space translation $\vec{v}$ (represented by a $d$-dimensional vector). For a spatial vector $\vec{r}$,
\begin{align}
\hat{g}\vec{r}=\{R|\vec{v}\} \vec{r}=R \vec{r}+\vec{v}.
\end{align}
The product of two group element is defined as
\begin{align}
\hat{g}\hat{g'}=\{R|\vec{v}\} \{R'|\vec{v}'\}=\{R R'|\vec{v}+R \vec{v}'\},
\label{eq:product}
\end{align}
and the identity operator of the group is
\begin{align}
\hat{e}=\{I|\vec{0}\},
\end{align}
where $I$ is the identity matrix and $\vec{0}$ is the vector with zero length. 

The elements of the space group with $\vec{v}=\vec{0}$ forms a subgroup of $G$, which is the point group of the lattice. Another important subgroup is formed by the elements with $R=I$, which gives all the space
translations commensurate with the lattice vector. By definition, equation~\eqref{eq:product} implies
that the space group is the semidirect product of the point group and the lattice translational group.

Consider the Bloch wavefunction $\Psi^{n}_{\vec{k}}(\vec{r})$, where $n$ is the band index
and $\vec{k}$ is the momentum. Here we drop the  spin index, which is irrelevant for our
discussion. Under the transformation $\hat{g}=\{R|\vec{v}\}$, the Bloch wavefunction at momentum $\vec{k}$ transfers into the Bloch wavefunction at momentum $R\vec{k}$:
\begin{align}
\{R|\vec{v}\}: \Psi^{n}_{\vec{k}}(\vec{r})
\rightarrow \sum_m U_{\hat{g},\vec{k}}^{nm} \Psi^{m}_{R \vec{k}}(\vec{r}),
\label{eq:Bloch_transformation}
\end{align}
where $U_{\hat{g},\vec{k}}$ is a unitary matrix.

For each momentum $\vec{k}$, the rotations which keep $\vec{k}$ invariant 
($R\vec{k}=\vec{k}$) form a subgroup of the point group of the lattice. This
subgroup is referred to as the little group $G_{\vec{k}}$. For the little group $G_{\vec{k}}$, 
equation~\eqref{eq:Bloch_transformation} reduces to
\begin{align}
R \Psi^n_{\vec{k}}(\vec{r})
=\sum_m U_{R,\vec{k}}^{nm}\Psi^m_{\vec{k}}(\vec{r}).
\end{align}
Here the unitary matrices $U_{R,\vec{k}}$ form
an unitary representation of the little group $G_{\vec{k}}$. 

In a square lattice, the point group is D$_4$. For 
the $\Gamma$ point [$\vec{k}=(0,0)$] and the $M$ point [$\vec{k}=(\pi/a,\pi/a)$], 
the little group coincides with the point group D$_4$. 
Therefore, at any of these two momentum points, we can repeat the same procedure as discussed 
in Sec. \ref{sec:single_site}. Similar conclusions are found concerning the phenomenon of 
the degeneracy. Namely, at the $\Gamma$ or $M$ point, if any of the Bloch waves  is odd under 
space-inversion, 
\begin{align}
\Psi^n_{\vec{k}}(\vec{r})=-\Psi^n_{\vec{k}}(-\vec{r}),
\end{align}
the states must be two-fold degenerate (i.e. two bands must cross at this momentum point).
On the contrary, if the Bloch waves are even in parity,
\begin{align}
\Psi^n_{\vec{k}}(\vec{r})=\Psi^n_{\vec{k}}(-\vec{r}),
\end{align}
they are in general nondegenerate.

\section{Degeneracy and parity of orbitals}
\label{sec:Wannier_function}
As shown above, the band degeneracy at $\Gamma$ and $M$ are determined by the parity of the Bloch
wavefunctions under space inversion. Here we show that this parity can be deduced from the parity of the orbitals which form the energy band.

In the tight-binding picture, the orbitals at each site forms energy bands as tunnelings are introduced.
Along this line of thinking, the Bloch wavefunction  $\psi^n_{\vec{k}}(\vec{r})$ 
can be written as a superposition of local 
orbitals as:
\begin{align}
\psi^n_{\vec{k}}(\vec{r})=\frac{1}{\sqrt{N}}\sum_{\vec{R},n} 
c_{n,m} \phi^m(\vec{r}) e^{i \vec{k}\cdot\vec{R}},
\label{eq:orbital_bloch}
\end{align}
where $n$ is the band index and $m$ indicates the orbitals on a single lattice site. $\phi^m(\vec{r})$ is the wavefunction for orbital $m$ at certain site and $c_{n,m}$ is a complex coefficient. $N$ in the normalization factor is the total number of sites of the lattice and $\vec{R}$ gives the coordinate of the sites of the Bravais lattice. 
 
Under the space inversion, $I$, we have
\begin{align}
\vec{R}\rightarrow -\vec{R},\\
\vec{r}\rightarrow -\vec{r},\\
\vec{k}\rightarrow -\vec{k}.
\end{align}
At $\Gamma$ and $M$, $e^{i \vec{k}\cdot\vec{R}}=\pm1$ and thus 
is invariant. Therefore, as can be seen from
equation~\eqref{eq:orbital_bloch}, the parity of a Bloch wavefunction 
at these two points is dictated by the parity of the orbitals that form this Bloch wave.
In the particular case studied in the main text, the band formed by space-inversion odd orbitals
are doubly degenerate at $\Gamma$ and $M$. This will be demonstrated in details
using a specific tight-binding model in Sec.~\ref{sec:tight_binding}

\section{Hybridization between orbitals of opposite parity}
\label{sec:tight_binding}
\subsection{Tight-binding model}

\begin{figure}
\begin{center}
\subfigure[]{\includegraphics[width=0.3\textwidth]{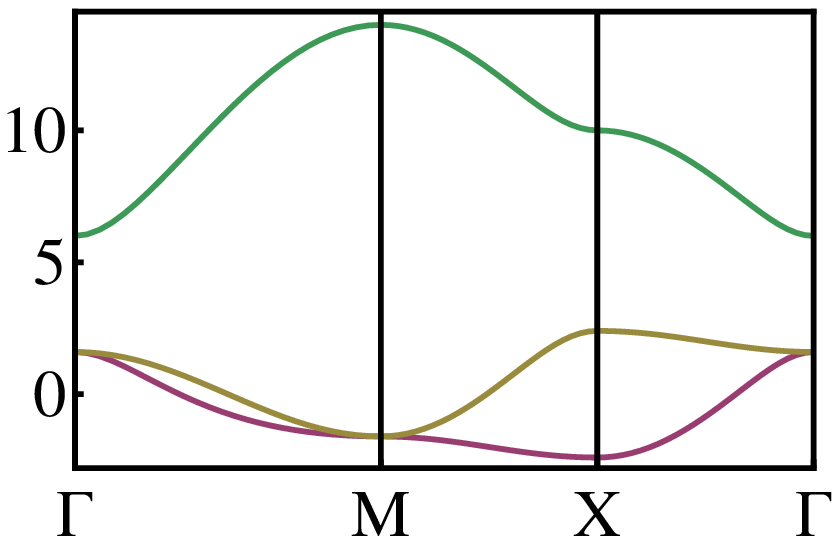}}
\subfigure[]{\includegraphics[width=0.3\textwidth]{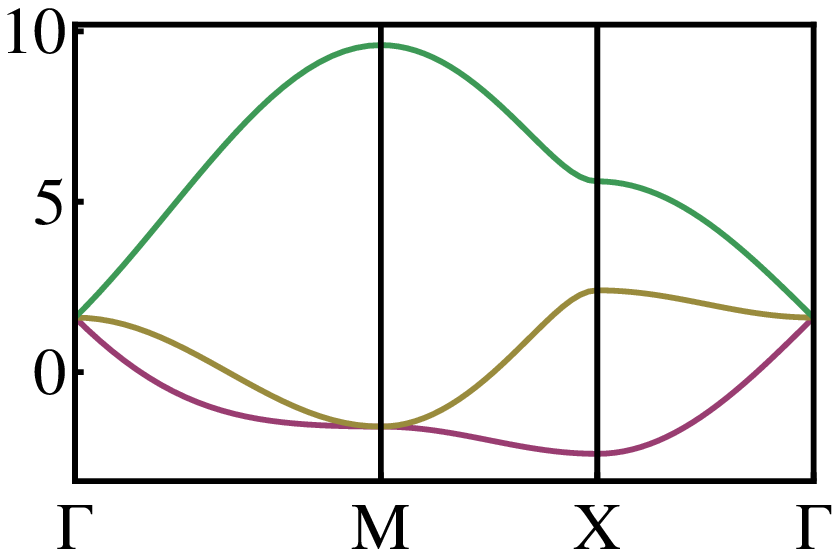}}
\subfigure[]{\includegraphics[width=0.3\textwidth]{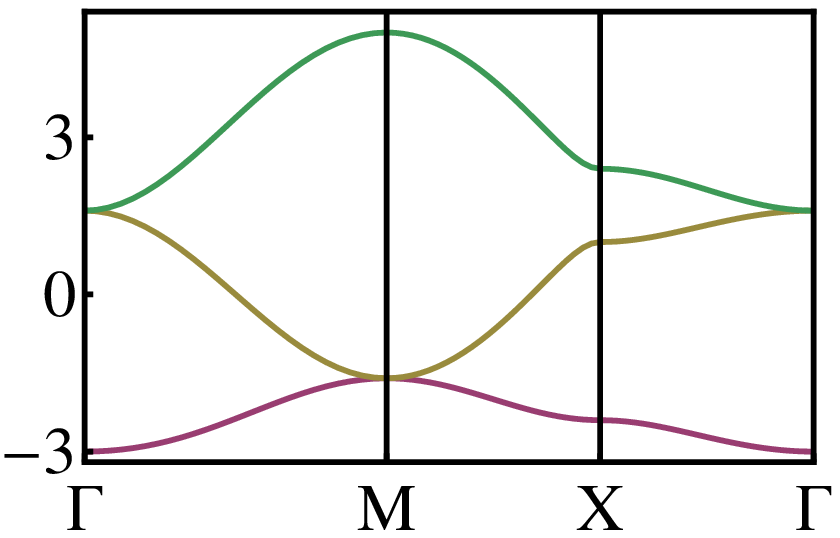}}
\end{center}
\caption{The single-particle energy spectrum in the tight-bind model.
The band structure of (a) is the
weak-hybridization case, while (c) shows the strong-hybridization regime. 
The marginal case is shown in  (b).
Here, we choose the hopping strength $t_{dd}=t_{pp}=t_{pd}=1$ and $t'_{pp}=0.2$.
From (a) to (c), the values of $\delta$ are $10$, $5.6$ and $1$. The tight-binding model
captures qualitatively the band structure  of the second, third and fourth band shown in 
Fig. \ref{fig:bands}.}
\label{fig:tight_binding_band}
\end{figure}

Considering a square lattice with three orbitals on each site ($p_x$, $p_y$ and $d_{x^2-y^2}$).
The Hamiltonian of the tight-binding model takes the following form
\begin{align}
H_0=&-t_{dd} \sum_{\vec{r}} (d_{\vec{r}}^\dagger d_{\vec{r}+\vec{a}_x}
+d_{\vec{r}}^\dagger d_{\vec{r}+\vec{a}_y}+h.c.)
\nonumber\\
&
+t_{pp} \sum_{\vec{r}}(p_{x, \vec{r}+\vec{a}_x}^\dagger p_{x,\vec{r}}
+p_{y,\vec{r}+\vec{a}_y}^\dagger p_{y,\vec{r}}+h.c.)
-t'_{pp} \sum_{\vec{r}} (p_{x,\vec{r}+\vec{a}_y}^\dagger p_{x,\vec{r}}
+p_{y, \vec{r}+\vec{a}_x}^\dagger p_{y,\vec{r}}+h.c.)
\nonumber\\
&+t_{pd} \sum_{\vec{r}} (p_{x, \vec{r}+\vec{a}_x}^\dagger d_{\vec{r}}
-p_{x, \vec{r}}^\dagger d_{\vec{r}+\vec{a}_x}
+p_{y,\vec{r}+\vec{a}_y}^\dagger d_{\vec{r}}
-p_{y,\vec{r}}^\dagger d_{\vec{r}+\vec{a}_y}
+h.c.)
\nonumber\\
&+\delta \sum_{\vec{r}} d_{\vec{r}}^\dagger d_{\vec{r}},
\label{eq:H_tight_binding}
\end{align}
where $\vec{r}$ is the coordinates of the lattice sites; $\vec{a}_{x}$ ($\vec{a}_{y}$) 
is the lattice vector in the $x$ ($y$) direction, and $p_{x,\vec{r}}$, $p_{y,\vec{r}}$ and $d_{\vec{r}}$ are the fermion annihilation operators for the $p_x$, $p_y$ and $d_{x^2-y^2}$ orbitals at site $\vec{r}$.
The hopping between orbitals on neighboring sites is described by the hopping amplitudes
$t_{dd}$, $t_{pp}$, $t_{pd}$ and $t'_{pp}$. The sign and strength of the hoppings are determined by the
over-lap between orbitals. In our convention, all the hopping strengths here are positive and $t'_{pp}$ is the smallest.
The last term gives the energy difference between $p$ and $d$ orbitals for a single site problem $\delta$,
which we assume to be positive.

In the momentum space, the tight-bind Hamiltonian becomes
\begin{align}
H_0=\sum_{\vec{k}}
\left(
\begin{array}{c c c}
d_{\vec{k}}^\dagger, & p_{x,\vec{k}}^\dagger, & p_{y,\vec{k}}^\dagger
\end{array}
 \right)
\mathcal{H}
\left(
\begin{array}{c}
d_{\vec{k}} \\
p_{x,\vec{k}}\\
p_{y,\vec{k}} \\
\end{array}
\right),
\label{eq:H_momentum_0}
\end{align}
where $d_{\vec{k}}$, $p_{x,\vec{k}}$ and $p_{y,\vec{k}}$ are the fermion annihilation 
operations at momentum $\vec{k}$  and
\begin{align}
\mathcal{H}=
\left(
\begin{array}{c c c}
-2 t_{dd} (\cos k_x+\cos k_y)+\delta & 2 i t_{pd} \sin k_x& 2 i t_{pd} \sin k_y \\
-2 i t_{pd} \sin k_x &2 t_{pp} \cos k_x-2 t'_{pp} \cos k_y & 0\\
-2 i t_{pd} \sin k_x &0 & 2 t_{pp} \cos k_y-2 t'_{pp} \cos k_x  \\
\end{array}
\right).
\label{eq:H_momentum}
\end{align}
For simplicity, we choose the lattice constant $a=1$. 

The eigenvalues of the $3\times3$ matrix in equation~\eqref{eq:H_momentum} gives the band structure. Depending on the value of $\delta$ (compared with the hopping strength), two different
types of band structures are found as shown in Fig.~\ref{fig:tight_binding_band}. For 
$\delta>4 t_{dd}+2 t_{pp}-2 t'_{pp}$, the weak hybridization is observed, but for $\delta<4 t_{dd}+2 t_{pp}-2 t'_{pp}$ the strong-hybridization regime is reached.
For the marginal case at  $\delta=4 t_{dd}+2 t_{pp}-2 t'_{pp}$, the three band touch at  $\Gamma$. 
This recovers the band structure calculated from plan wave expansion for the optical lattice presented in the main text, if we identify these three bands as the second, third and fourth band there.

\subsection{Generalization}
It turns out that the conclusion above can be generalized to other orbitals of opposite parity. 
To demonstrate this, we consider a general 
tight-binding model on a square lattice. We further assume that at each site, only one parity even orbital and a pair of parity odd orbitals are close to the chemical potential, so that we can ignore all other orbitals in the leading order approximation. Here we consider two parity odd orbtials because for any potential wells with four-fold rotational symmetry, the odd parity orbitals are doubly degenerate, while the even parity orbitals are in general nondegenerate.
We also assume the energy difference between of the parity-even and parity-odd orbitals is 
described by a parameter $\delta$. 

When $\delta$ is much larger than the hopping strength, the parity-even and parity-odd 
orbitals are separated in energy. Therefore, the mixing between these two types of orbitals is 
small. Roughly speaking, in this limit, the parity-even orbitals will form one band, 
which in general has no degeneracy point. The parity-odd orbitals form two bands, 
which cross each other at $\Gamma$ and $M$, due to the odd parity of the orbitals. 
This is the weak strong-hybridization limit.

On the other hand, when $\delta$ is small compared with the hopping strength,
the hybridization between parity-even and parity-odd becomes important. 
In this case, it is no longer possible to identify the distinct parity-even band and 
parity-odd bands, since all the three bands are now mixed together. Here, 
the upper two bands cross at $\Gamma$ (or $M$) and the lower two bands cross 
at $M$ (or $\Gamma$), which gives us the strong-hybridization limit.

\section{Topological protection of the band degeneracy points}
\label{sec:S7}
In this section we provide two equivalent topological indices to prove the topological nature 
of the band degeneracy. The first approach shows that
the band degeneracy point can be considered as a vortex in the momentum space with integer winding numbers, and the second one presents the same topological structure through the concept of the Berry flux \cite{Blount1962, Haldane2004}.



\subsection{Band degeneracy as a topological defect in momentum space}
\label{sec:winding_number}

Consider the tight-binding model presented in equation~\eqref{eq:H_momentum_0}.
Near the band degeneracy point at $\Gamma$, it is easy to check that the energy of the
fermion modes from $d$ orbitals are far away from the chemical potential, 
so we can integrate out these high-energy degrees of freedom and focus only on the low-energy 
modes (from $p$ orbitals). By doing so, the Hamiltonian is reduced to an 
effective two-band model with
\begin{align}
H_0=\sum_{\vec{k}}
\left(
\begin{array}{c c}
p_{x,\vec{k}}^\dagger & p_{y,\vec{k}}^\dagger
\end{array}
 \right)
\tilde{\mathcal{H}}
\left(
\begin{array}{c}
p_{x,\vec{k}}\\
p_{y,\vec{k}} \\
\end{array}
\right).
\label{eq:Hamiltonian_reduced_0}
\end{align}
Here, although the $d$ orbitals drop off from the Hamiltonian,
the kernel of the Hamiltonian, $\tilde{\mathcal{H}}$,
receives corrections from virtual processes
in which a particle jumps from a $p$ orbital to a $d$ orbital and then back 
to a $p$ orbital (and other higher-order virtual processes, which we would not consider here). 
Using perturbation theory, 
the matrix  $\tilde{\mathcal{H}}$ can be determined order-by-order as
\begin{align}
\tilde{\mathcal{H}}
=\left(\begin{array}{c c}
\mathcal{H}_{22} & \mathcal{H}_{23}\\
\mathcal{H}_{32} & \mathcal{H}_{33}\\
\end{array}\right)
-
\frac{1}{H_{11}-\mu}\left(\begin{array}{c c}
\mathcal{H}_{21}\mathcal{H}_{12} 
& \mathcal{H}_{21}\mathcal{H}_{13}
\\
\mathcal{H}_{31}\mathcal{H}_{12} 
& \mathcal{H}_{31}\mathcal{H}_{13}
\end{array}\right)
+\ldots
\end{align}
where $\mu$ is the chemical potential and $\mathcal{H}_{ij}$ is the $(i,j)$ component
of the matrix shown in equation~\eqref{eq:H_momentum}.
Here, the first term on the right hand side is the zeroth order term in the perturbation
expansion, generated by direct hoppings between $p$ orbitals, 
while the second term is from the second order perturbation, which describes the virtual hopping processes mentioned above.

By expanding the momentum around the $\Gamma$ point, to the leading order,
equation~\eqref{eq:Hamiltonian_reduced_0} becomes
\begin{align}
H_0=\sum_{\vec{k}}
\left(
\begin{array}{c c}
p_{x,\vec{k}}^\dagger & p_{y,\vec{k}}^\dagger
\end{array}
 \right)
\left(
\begin{array}{c c}
 t_1 k_x^2+t_2 k_y^2 &  2t_3 k_x k_y\\
2t_3 k_x k_y & t_1 k_y^2+t_2 k_x^2 
\end{array}
\right)
\left(
\begin{array}{c}
p_{x,\vec{k}}\\
p_{y,\vec{k}} \\
\end{array}
\right),
\label{eq:Hamiltonian_reduced}
\end{align}
where
\begin{align}
t_1&= t_{pp}+\frac{4 t_{pd}^2}{2 t_{pp}-2t_{pp}'+4 t_{dd}-\delta},
\\
t_2&=-t_{pp}',
\\
t_3&=\frac{2 t_{pd}^2}{2 t_{pp}-2t_{pp}'+4 t_{dd}-\delta}.
\end{align}
This Hamiltonian recovers the general model of a quadratic-band crossing point 
studied in Ref.~\cite{sun2009}.

\begin{figure}
\begin{center}
\subfigure[]{\includegraphics[width=0.3\textwidth]{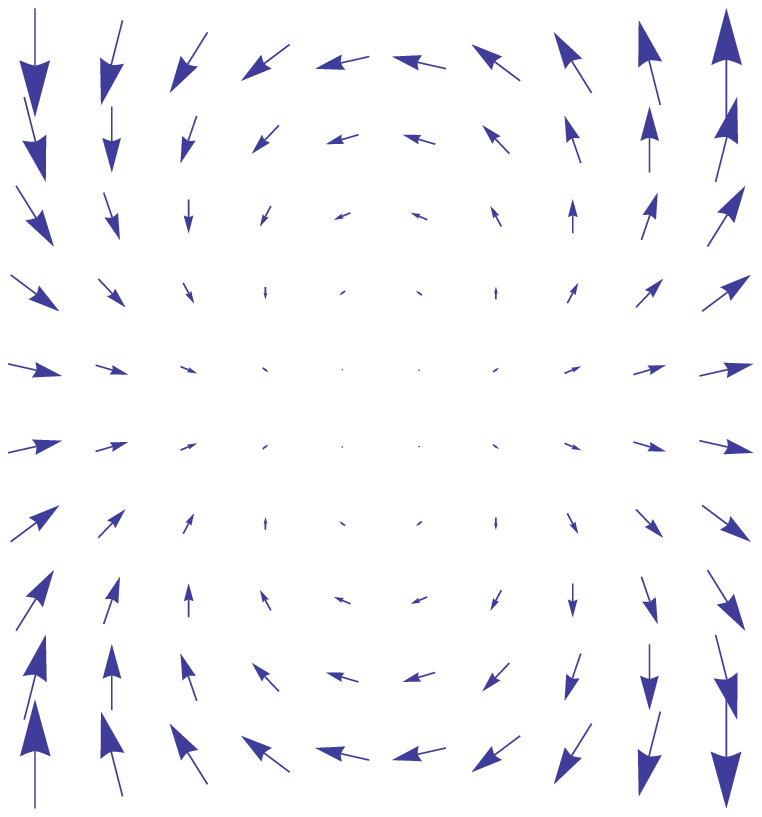}}
\subfigure[]{\includegraphics[width=0.3\textwidth]{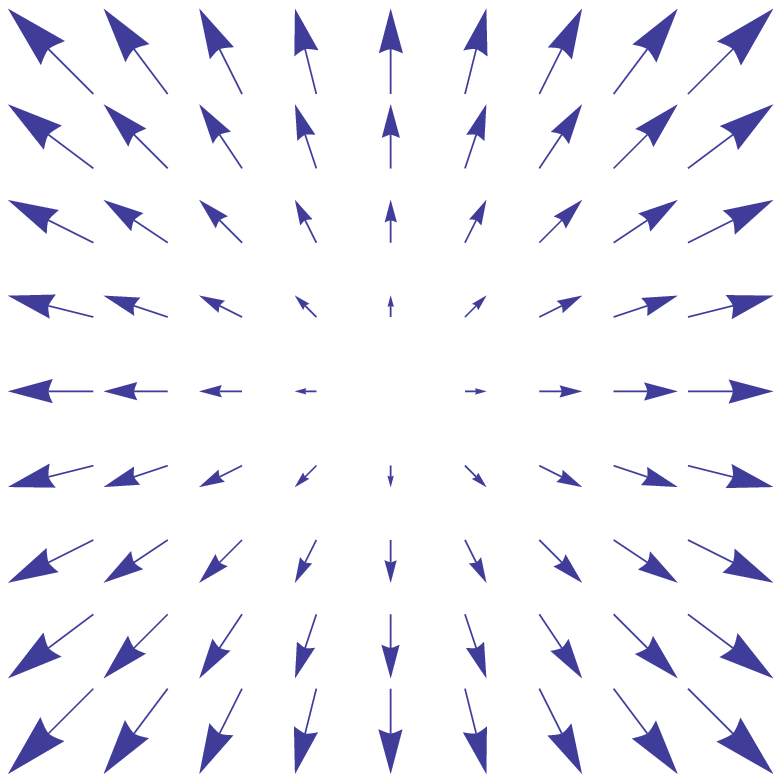}}
\end{center}
\caption{Band degeneracy point as a topological defect. Here (a) shows the planar vector $\vec{h}$
defined in equation~\eqref{eq:vector} in the momentum space near the band degeneracy point that we found in the square lattice model. For comparison, we show in (b) the same planar vector near
the Dirac point as in graphene.
In (c), we show that the band degeneracy point we found actually
corresponds to a vortex in the momentum space with winding number $2$. Here the vector field
$\mathcal{H}$ is defined in equation~\eqref{eq:vector}.
For comparison, we also show the same vector field for a Dirac point, which corresponds to a topological
defect with winding number $1$. This nontrivial winding number indicates that the band degeneracy is
topologically protected and thus can not be lifted in adiabatic procedures, unless the symmetry of the system is broken.}
\label{fig:vortex}
\end{figure}

The $2\times 2$ matrix in equation~\eqref{eq:Hamiltonian_reduced} 
can be expanded in the basis of four  independent matrices of SU(2) algebra as follows
\begin{align}
\mathcal{H}=\frac{t_1+t_2}{2} (k_x^2+k_y^2) I+2 t_3 k_x k_y \sigma_x
+\frac{t_1-t_2}{2}(k_x^2-k_y^2) \sigma_z,
\label{eq:pauli}
\end{align}
where $\sigma_x$ and $\sigma_z$ are the two Pauli matrices and $I$ is the $2\times 2$ identity
matrix. We can define a 2D planar vector using the coefficients of the two Pauli matrices
\begin{align}
\vec{h}=\left(2 t_3 k_x k_y, \frac{t_1-t_2}{2}(k_x^2-k_y^2)\right).
\label{eq:vector}
\end{align}
The length of this vector $\left|\vec{h}\right|$ has the physical meaning of (half of) the band gap between 
the two bands, if one notice that the dispersion relation for the two bands are
\begin{align}
E_{\pm}=\frac{t_1+t_2}{2} (k_x^2+k_y^2) \pm |\vec{h}|.
\end{align}

As shown in Fig.~\ref{fig:vortex}, this 2D planar vector $\vec{h}$ has a vortex structure
at $\vec{k}=0$. A vortex is a topological defect described by the topological index (the winding number), 
\begin{align}
W=\oint_\mathcal{C} \frac{d \vec{k}}{2\pi} \cdot \left[\frac{h_x}{|\vec{h}|} 
\vec{\nabla}\left(\frac{h_y}{|\vec{h}|}\right)-\frac{h_y}{|\vec{h}|} 
\vec{\nabla}\left(\frac{h_x}{|\vec{h}|}\right)\right].
\end{align}
For the topological semimetal which we consider here, the winding number is $W=2$. For comparison, the vortex structure for a Dirac point is also plotted, which has winding number $1$.

At the vortex core, $\vec{h}=0$, which indicates a band degeneracy point. Thus, the band degeneracy can be considered as a topological defect in the momentum space. It is this topological nature that protects 
the band degeneracy that we have found from being lifted against any adiabatic deformation of the Hamiltonian.

In fact, this vortex structure is generic for symmetry protected band-degeneracy point. In general,
if we focus on the two bands of the degeneracy point 
and treat the system as a two-band model, the Hamiltonian can 
always be expanded into the identity matrix and the Pauli matrices, as in equation~\eqref{eq:pauli}. In the presence of the time-reversal and space inversion symmetries, at most two Pauli matrices can appear while the third one is forbidden due to symmetry.
Thus, we can use the coefficients of the two Pauli matrices to form a 2D planar vector as in 
equation~\eqref{eq:vector} and the length of this vector gives the band gap between the two bands.

\subsection{Berry flux and degeneracy points}
\label{sec:Berry_flux}
The same topological nature can be presented in terms of the Berry flux. The Berry flux is 
defined as the contour integral of the Berry connection in the momentum space~\cite{Blount1962, Haldane2004}.
\begin{align}
\Phi_B^n=\oint_{\mathcal{C}} d \vec{k} \cdot \vec{\mathcal{A}}^n_{\vec{k}},
\end{align}
where $n$ is the band index and $\mathcal{C}$ is certain closed contour in the momentum space. Here, the Berry connection $ \vec{\mathcal{A}}^n_{\vec{k}}$ is defined as 
\begin{align}
 \vec{\mathcal{A}}^n_{\vec{k}}=-i \int d^2 r 
\Psi^n_{\vec{k}}(\vec{r})^* \vec{\nabla}_{\vec{k}} \Psi^n_{\vec{k}}(\vec{r}),
\end{align}
with $\Psi^n_{\vec{k}}(\vec{r})$ being the Bloch wave of band $n$ at momentum $\vec{k}$. The Berry connection $\vec{\mathcal{A}}^n_{\vec{k}}$ can be roughly considered as a gauge field in the momentum space. In this analogy, the Berry flux corresponds to the magnetic flux in the area enclosed by contour $\mathcal{C}$. Similar as the flux quantization of the magnetic field, the Berry flux of a time-reversal invariant system is quantized to $m \pi$ with $m$ being an integer~\cite{Blount1962, Haldane2004}.

However, in the definition of the Bloch wavefunctions, there is an undetermined $U(1)$ phase that
can be chosen arbitrarily. This can be seen by noticing that for any Bloch wave 
$\Psi^n_{\vec{k}}(\vec{r})$, the following equation,
\begin{align}
\tilde{\Psi}^n_{\vec{k}}(\vec{r})=e^{i \phi(\vec{k})}\Psi^n_{\vec{k}}(\vec{r}),
\label{eq:U1_phase}
\end{align}
also defines a Bloch wave where $\phi(\vec{k})$ is an arbitrary function of momentum $\vec{k}$.
For any $\phi(\vec{k})$, the integral over a closed contour $\mathcal{C}$ in the BZ,
\begin{align}
W=\oint_{\mathcal{C}} d\vec{k} \nabla_{\vec{k}}\phi(\vec{k}),
\label{eq:loop_phase}
\end{align}
always leads to an integer value  due to the periodical structure
around the contour.

For a general contour $\mathcal{C}$, the redefinition of phase in equation~\eqref{eq:U1_phase} changes 
the value of the Berry flux by $2 W \pi $. Therefore, the Berry flux is only well defined up to mod 
$2\pi$. Since the Berry flux is quantized to integer multiplied by $\pi$, only two classes of
flux, $\Phi_B=0$ and $\pi$, can be distinguished. All other values of $\Phi_B$ can be connected to these 
two classes via redefinition of phase in equation~\eqref{eq:U1_phase}.

However, for a system with space inversion symmetry, if the contour was also chosen in such a way that it is invariant under space inversion (i.e. if $\vec{k}$ is a point on the contour, so is $-\vec{k}$), then the Berry flux can be defined up to mod $4\pi$. This can be achieved by requiring
\begin{align}
I \Psi^n_{\vec{k}}(\vec{r})=\Psi^n_{-\vec{k}}(\vec{r}),
\end{align}
where $I$ is the space-inversion operator. This is alway possible by choosing a proper $\phi$ and
redefine the Bloch waves as in equation~\eqref{eq:U1_phase}. With this constraint, the redefinition of phase 
[equation~\eqref{eq:U1_phase}] can only be allowed for $\phi(\vec{k})=2n\pi+\phi(-\vec{k})$ with $n$ 
being an integer. Therefore, the winding number $W$ [equation~\eqref{eq:loop_phase}] must be even, 
$W=2 m$ with $m$ being an integer. 
Thus, the Berry flux will now change by $4 m \pi$ under any phase redefinition. As a result, the
Berry flux becomes well defined up to mod $4 \pi$. Opposite to the general case discussed above, now the Berry flux $2\pi$ and $0$ are topologically distinguished states and cannot be adiabatically deformed into each other. This is one example where a discrete symmetry of the system (e.g. the point group symmetry) changes the topological classification of a system.

For the problem we study here, we can always choose the eigenvector 
[$a_{n,m}$ in equation~\eqref{eq:plan_wave_expansion}] to be real. 
This enables the computation of the Berry flux via a simpler method. 
If one considers the phase uncertainty shown in equation~\eqref{eq:U1_phase} as a gauge-like
symmetry in the momentum space, this approach is in analogy to adopting a specific gauge which
helps simplify the computation for physical quantities.
We first require all the coefficients
$a_{n,m}$ to be real. However, there is still an arbitrary sign undetermined for $a_{n,m}$ 
at each $\vec{k}$. To fix this sign, we choose a contour around the $\Gamma$
or $M$ point and require the wavefunction to satisfy the relation
\begin{align}
\int  [\Psi^n_{\vec{k}}(\vec{r})]^*\Psi^n_{\vec{k}'}(\vec{r}) d^2 r=1+O(|\vec{k}-\vec{k}'|),
\end{align}
for any two momentums $\vec{k}$ and $\vec{k}'$ on the contour which are close to each other.
Now, one still has the freedom to flip the sign for $a_{n,m}$ for 
all the points on the contour at the same time, but up to this sign, the value of $a_{n,m}$ is determined,
which can be viewed as a gauge fixing.

Due to the space inversion symmetry,  the coefficients $a_{n,m}$ at $\vec{k}$ and $-\vec{k}$
must have the same amplitude but can have either the same or opposite signs:
\begin{align}
a_{n,m}(\vec{k}) = a_{n,m}(-\vec{k}), \textrm{   or   } 
a_{n,m}(\vec{k}) =-a_{n,m}(-\vec{k}).
\end{align}
The Berry flux is zero for the first case but $2\pi$ for the second.

In the definition of the Berry flux, one uses the Bloch wavefunction in a specific band, say $n$. Therefore,
the contour integral is well defined, when there is no band degeneracy on the contour.
On the one hand, if the area enclosed by the contour has no band degeneracy point inside, one can continuously 
deform (shrink) the contour into a point, which by definition has zero Berry flux. During this adiabatic deformation of the contour, every quantity should change in a continuous way. On the other hand, the Berry flux is quantized and cannot be tuned adiabatically. Therefore, the Berry flux must be invariant during this process.These arguments lead to the following conclusion: for any contours enclosing no band degeneracy points for the band $n$, the Berry flux of this band must be zero.
Now to the contrary, if the Berry flux is nonzero, this contour must not be able to shrink adiabatically to a point. The only way this is possible is when there is a band degeneracy point enclosed by this contour. Thus we prove that a nonzero Berry flux is the sufficient condition to have a protected band degeneracte point.

We end the discussion about the topological nature of the band degeneracy by unifying the two 
topological indices that we have computed in this section. In fact, if one multiplies the winding number studied in 
Sec.~\ref{sec:winding_number} by $\pi$, it coincides with the Berry flux examined in Sec.~\ref{sec:Berry_flux}.

\section{Instability induced by repulsive interactions}
The low-energy effective theory we constructed in Sec.~\ref{sec:winding_number} [equation~\eqref{eq:Hamiltonian_reduced}] can be mapped exactly to
the general theory studied in Refs.~\cite{sun2009,sun2008}. By mapping the results found in Ref.~\cite{sun2009} back into our model, the interaction effect can be solved in the weak-coupling limit.
Here we summarize the results from this mapping and provide a simple mean-field picture for understanding this results.

In the low-energy effective theory for fermions with two bands (but no spin),
only one interaction term needs considering,
\begin{align}
H_{\textrm{int}}= V \sum_{\vec{r}} p_{x,\vec{r}}^\dagger p_{x,\vec{r}}
p_{y,\vec{r}}^\dagger p_{y,\vec{r}}.
\label{eq:interaction}
\end{align}
Other (short-range) interaction terms involve derivatives and thus are irrelevant in the sense of renormalization group (RG).  As shown in Ref. \cite{sun2009}, at low temperature, this interaction is marginal at tree level under RG, while the one-loop RG calculation shows that this interaction is marginally relevant for $V>0$ (repulsive) and marginally irrelevant with $V<0$. This implies that when the temperature is lowered for this system, for particles with attractive interactions, the noninteracting band structure theory remains accurate in the low energy limit. However, if the interaction is repulsive, it leads to an instability once the temperature is below certain critical value $T_C$. It worth emphasizing that the scaling behavior here is very similar to the BCS theory in Fermi liquids, except that the instability shows opposite behavior regarding the sign of interaction. In the latter case, the system is unstable for attractive interactions, but remains stable for repulsive interactions.

The result of this instability is a state with spontaneously generated angular momentum. This can
be seen if one notices that the interaction term in equation~\eqref{eq:interaction} can be reformulated as
\begin{align}
H_{\textrm{int}}=-\frac{V}{2}\sum_{\vec{r}} (L^z_{\vec{r}})^2,
\label{eq:interaction2}
\end{align}
up to some unimportant constant 
where 
\begin{align}
L^z_{\vec{r}}=-i (p_{x,\vec{r}}^\dagger p_{y,\vec{r}}-p_{y,\vec{r}}^\dagger p_{x,\vec{r}}),
\end{align}
is the angular momentum in the $z$-direction at site $\vec{r}$. The physical meaning of 
$L^z_{\vec{r}}$ can be checked by noticing
\begin{align}
L^z |p_x+i p_y\rangle=+|p_x+i p_y\rangle,
\\
L^z |p_x-i p_y\rangle= -|p_x-i p_y\rangle.
\end{align}
Here we verify equation~\eqref{eq:interaction2} by expanding it using the fermion creation 
and annihilation operators:
\begin{align}
H_{\textrm{int}}=&-\frac{V}{2}\sum_{\vec{r}} (L^z_{\vec{r}})^2
=\frac{V}{2}\sum_{\vec{r}}(p_{x,\vec{r}}^\dagger p_{y,\vec{r}}p_{x,\vec{r}}^\dagger p_{y,\vec{r}}-p_{x,\vec{r}}^\dagger p_{y,\vec{r}}p_{y,\vec{r}}^\dagger p_{x,\vec{r}}-p_{y,\vec{r}}^\dagger p_{x,\vec{r}}p_{x,\vec{r}}^\dagger p_{y,\vec{r}}+p_{y,\vec{r}}^\dagger p_{x,\vec{r}}p_{y,\vec{r}}^\dagger p_{x,\vec{r}}).
\end{align}
Notice that the first and last term on the right hand side vanish because 
$p_{x,\vec{r}}^\dagger p_{x,\vec{r}}^\dagger=p_{y,\vec{r}}^\dagger p_{y,\vec{r}}^\dagger=0$.
The two remaining terms recover exactly the equation~\eqref{eq:interaction}, with an extra constant
term: $V N/2$ where N is the total particle number in the system.

From equation~\eqref{eq:interaction2}, it is transparent that repulsive interaction favors $L^z\ne 0$,
Since the repulsive interaction is marginally relevant, it shall dominate the low energy behavior
of the system. Therefore, repulsive interaction shall result in the low-temperature instability towards a state with spontaneously generated angular momentum and $\avg{L^z}$ is the order parameter for this phase.
The nonezero value of this order parameter indicates that the low-temperature ordered phase breaks
spontaneously the time-reversal symmetry and also reduces the point group symmetry from D$_4$ down
to C$_2$. 
Due to this symmetry breaking, the band degeneracy is no longer symmetry protected, and a gap
opens at the band  point. As a result, the system turns into an insulator with a finite 
energy gap. This insulator has a nontrivial topological structure, with the first Chern number being $1$. The topological properties of this state is the same as the quantum Hall state with filling $1$, but it is induced by interaction and spontaneous symmetry breaking, instead of a magnetic field. 
An insulating state of this kind of topological properties is often referred to as the 
quantum anomalous Hall state.~\cite{Haldane1988}

For larger $V$, although a Hartree-Fock treatment can still be preformed, the mean-field approximation becomes invalid as the energy scale of interaction becomes comparable with the band width. Furthermore, the low-energy effective theory, which we obtained by starting from a noninteracting band structure and treating interactions as perturbations,  is not expected to describe the properties of a system with strong interactions. In this limit, the behavior of the system becomes
sensitive to the microscopic details. 

We repeated the same mean-field calculation including the $d$-orbitals and all the conclusions above remains the same. As have been shown in Sec. \ref{sec:tight_binding} and \ref{sec:S7}, the energy of the $d$-orbitals are far from the Fermi level near the band degeneracy point and the degeneracy at $\Gamma$ is between the two p-bands. Therefore, the interactions between $d$ orbitals are not relevant for analyzing the instability of this level degeneracy point, and the same is true for interactions between $p$ and $d$ orbitals. The only relevant interaction term here is the repulsions between $p$ orbitals, same as observed above in the two-band approximation. 
In the full three-band model, this interactions can be decoupled using
the standard mean-field approximation as
\begin{align}
H_{MF}=H_0 - V \sum_{\vec{r}} \avg{L^z_{\vec{r}}}L^z_{\vec{r}}
+\frac{V}{2}\sum_{\vec{r}} (L^z_{\vec{r}})^2
\label{eq:H_MF}
\end{align}
where $H_0$ is the hopping part defined in Eq.~\eqref{eq:H_momentum} and $\avg{L^z_{\vec{r}}}$ is the order parameter which are determined by minimizing the free
energy. Same as the two-band model we computed above, this three-band model shows the same instability at infinitesimal $V>0$. We compared other competing orders, such as nematic and 
density waves, and found that the topological insulating state is preferred.
In the ordered phase with $\avg{L^z_{\vec{r}}}\ne 0$, we can use the mean-field Hamiltonian 
[Eq.~\eqref{eq:H_MF}] to compute the energy spectrum. By solving the energy spectrum on a cylindrical geometry, we observed
the topological edge modes as shown in Fig.~3 of the letter.

\section{Detecting the topological semimetal using Bragg scattering}

\begin{figure}
\begin{center}
\includegraphics[width=0.6\textwidth]{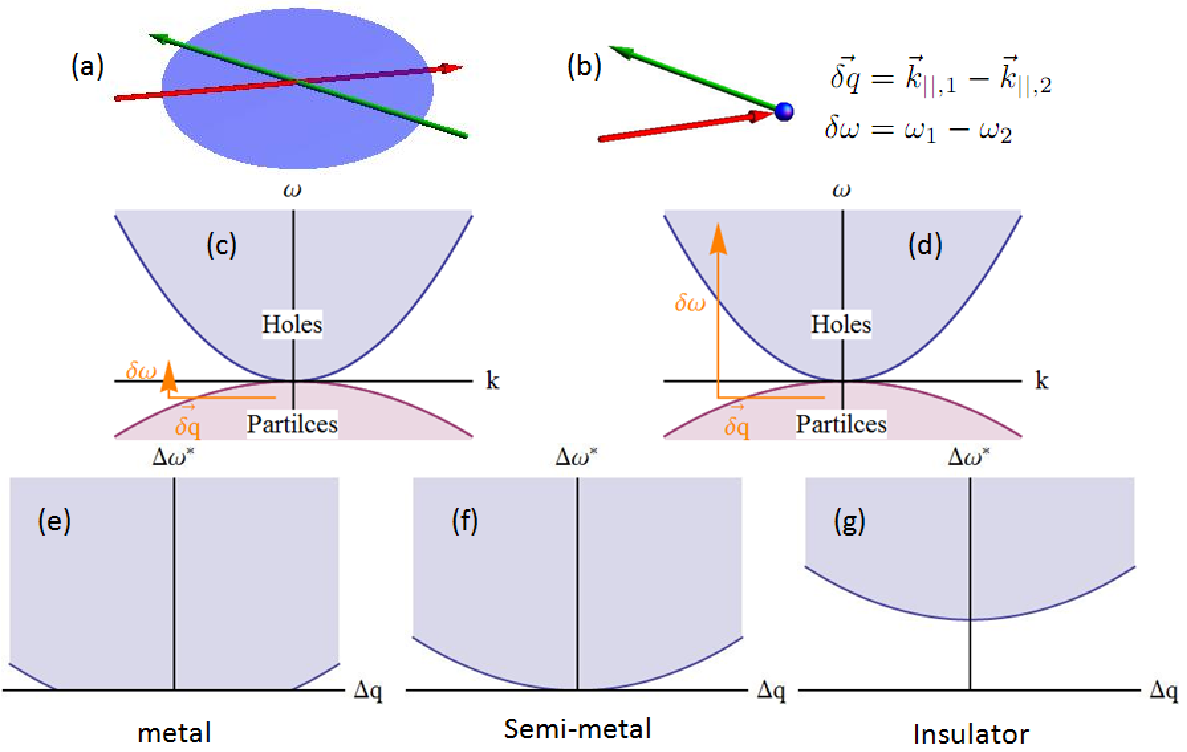}
\end{center}
\caption{Experimental signatures for the topological semimetal and topological insulator in Bragg scattering. On the top row, the experimental setup of Bragg scattering is shown in (a) and the schematic scattering 
process is shown in (b) along with the formula for $\vec{\delta{q}}$ and $\delta\omega$. (c) and (d) demonstrate the scattering processes with small and large $\delta \omega$ respectively. 
The former processes is forbidden, due to the lack of a final state which can satisfy the energy and momentum conservation laws. The threshold for $\delta \omega$ above which the scatterings are allowed is shown in Figs. (e)-(g) as a function of the momentum change $\delta q$, for a metal, topological semimetal and insulator accordingly. 
}
\label{fig:exp}
\end{figure}

Here, we briefly describe the experimental signatures of a semimetal state with respect to the application of Bragg scattering. 
When two laser beams with different frequencies and wavevectors are used, the particles in the system can absorb a photon from one beam and emit a photon into the other beam. 
In this process, the momentum gained by a particle confined in a 2D plane is $\vec{\delta q}=\vec{k}_{||,1}-\vec{k}_{||,2}$,
where $\vec{k}_{||,1}$ and $\vec{k}_{||,2}$ are the in-plan components of the wavevectors of
the two laser beams. In addition, the energy of the particle will change by $\delta\omega=\omega_1-\omega_2$ with $\omega_1$ and $\omega_2$ being the frequencies of the beams. While $\omega$ can be tuned by changing the frequency of the two laser beams, $\vec{\delta q}$ can be tuned independently via tilting the direction of the beams out of the 2D plane. Therefore, we can treat $\vec{\delta q}$ and $\delta \omega$ as two independent control parameters. At a given value of $\vec{\delta q}$, as shown in Fig.~\ref{fig:exp},  Bragg scattering is only allowed when $\delta \omega$ exceeds a threshold $\delta\omega^*$. The curves of $\delta\omega^*$ as a function of  $\vec{\delta q}$ carry direct information on the band structure. For the
topological semimetal, $\delta\omega^*$  is a quadratic function of  $\vec{\delta q}$ and 
vanishes at $\vec{\delta q}=0$. On the contrary,
$\delta\omega^*>0$ for an insulator at any $\vec{\delta q}$, while $\delta\omega^*=0$ for a range of $\vec{\delta q}$ in a metal.


\end{document}